\documentclass[%
 reprint,
superscriptaddress,
 amsmath,amssymb,
 aps,
pra,
]{revtex4-2}
\usepackage[colorlinks,allcolors=blue]{hyperref}

\usepackage{float}
\usepackage{xcolor}
\usepackage{graphicx}
\usepackage{dcolumn}
\usepackage{bm}
\usepackage{bbm}
\usepackage[normalem]{ulem}
\usepackage[T1]{fontenc}  
\usepackage{ulem}

\usepackage{amsmath}
\usepackage{braket}
\usepackage{dirtytalk}
\usepackage{multirow}
\usepackage[UKenglish]{babel}

\begin{document}

\preprint{APS/123-QED}

\title{Machine Learning-Based Characterisation of the Non-Markovian Dynamics of a Nitrogen-Vacancy Centre}

\author{T.\ Lannon}
\email{tlannon01@qub.ac.uk}
\affiliation{Centre for Quantum Materials and Technologies, School of Mathematics and Physics, Queen's University Belfast, BT7 1NN, United Kingdom}
\author{M.\ Paternostro}
\affiliation{Quantum Theory Group, Dipartimento di Fisica e Chimica Emilio Segrè, Università degli Studi di Palermo, via Archirafi 36, I-90123 Palermo, Italy}
\affiliation{Centre for Quantum Materials and Technologies, School of Mathematics and Physics, Queen's University Belfast, BT7 1NN, United Kingdom}
\author{J.M.\ Gregg}
\affiliation{Centre for Quantum Materials and Technologies, School of Mathematics and Physics, Queen's University Belfast, BT7 1NN, United Kingdom}
\author{A.\ Kumar}
\affiliation{Centre for Quantum Materials and Technologies, School of Mathematics and Physics, Queen's University Belfast, BT7 1NN, United Kingdom}

\date{\today}

\begin{abstract}
The interaction between a quantum system and its environment can be characterized by the spectral density function: knowing its structure is important for optimizing applications of quantum technologies such as quantum sensing protocols. In this work, we present the first experimental demonstration of a machine learning-based reconstruction of reaction-coordinate spectral density parameters from NV centre Rabi dynamics. Unlike the previous work, we recover all spectral density parameters rather than only the central frequency, and benchmark the performance of the neural network against the Cramér-Rao bound and maximum likelihood estimator. Our results demonstrate that the model predicted by the neural network can reliably reproduce the NV dynamics over the estimation window, and can produce estimates for some parameters with variances comparable to that of maximum likelihood.
\end{abstract}

\maketitle

\section{Introduction}
Understanding how a quantum system interacts with its surroundings is crucial for tasks such as quantum sensing and optimal control \cite{poggiali_optimal_2018}. Methods for characterising these surroundings typically involve using the qubit as a sensor of its own environment. By preparing the qubit in a known state and allowing it to interact with its environment for a given time, information about the structure of the environment becomes encoded in the state of the qubit \cite{degen_quantum_2017}. Measurements on the qubit and suitable post-processing of the results allow one to obtain information about the environment's structure.\par
A powerful method to describe the dynamics of open quantum systems is to derive a master equation from the spin-boson model. Here, the environment consists of an infinite bath of quantum harmonic oscillators coupled to the system of interest, with the distribution of coupling strengths determined by the spectral density function. In the limit of a continuum of modes, the spectral density can be approximated by a smooth function \cite{breuer_theory_2007,weiss_quantum_2008}, with the exact form dependent on the physical system in question. The parameters that specify the spectral density function determine the open system dynamics and hence developing methods to infer their values from experimental data has been a focus of recent research \cite{benedetti_characterization_2014,benedetti_quantum_2014,hernandez-gomez_noise_2018,martina_deep_2023,zwick_maximizing_2016,bina_continuous-variable_2018,benedetti_quantum_2018,tamascelli_quantum_2020,zhang_improving_2024,yuge_measurement_2011}. In cases where the spectral density produces weak coupling between the system and environment, the flow of information from the qubit to the environment is monotonic, permitting the Born and Markov approximations. However, some choices of spectral density invalidate these approximations \cite{breuer_colloquium_2016}. As such, there has been considerable effort to define, quantify,  and measure non-Markovianity in quantum systems \cite{rivas_quantum_2014}, due to the potential for information backflow to enhance, for example, quantum sensing capabilities \cite{https://doi.org/10.1002/qute.202400094}.\par
One approach to incorporating non-Markovian dynamics into the master equation description is the reaction coordinate (RC) mapping \cite{garg_effect_1985,thoss_self-consistent_2001,nazir_reaction_2018,martinazzo_communication_2011,strasberg_nonequilibrium_2016,correa_pushing_2019}. This constitutes a Bogolubov transformation of the environmental modes into a strongly coupled effective mode and a weakly coupled residual bath. This allows the system and reaction coordinate to be simulated exactly, while the Born and Markov approximations can be made to simulate the interaction with the residual bath. Within this formalism, the spectral density can be written as a sum of Lorentzian peaks, a form which often appears in solid-state systems \cite{norambuena_spin-lattice_2018,norambuena_quantifying_2020} and the field of quantum biology \cite{huelga_vibrations_2013,lambert_quantum_2013}. These structured spectral densities contain sharp peaks where environmental modes are more strongly coupled to the system, leading to non-Markovian effects.\par
Machine learning techniques have had success in a variety of applications in quantum information processing, including quantum tomography \cite{palmieri_experimental_2020, torlai_neural-network_2018,banchi_modelling_2018}, quantum control \cite{giannelli_tutorial_2022,niu_universal_2019,sgroi_reinforcement_2021,brown_reinforcement_2021,khalid_sample-efficient_2023,sgroi_reinforcement_2025}, quantum channel discrimination \cite{kardashin_quantum-machine-learning_2022}, and simulation and characterisation of open quantum system dynamics \cite{carleo_solving_2017,luo_autoregressive_2022,bandyopadhyay_applications_2018,luchnikov_simulation_2019,wei_finding_2024,nelson_data-driven_2022,goswami_experimental_2021,mahlow_predicting_2024}. In particular, neural networks have been applied to identify the central frequencies of a spectral density formed by the sum of one, two, and three Lorentzian functions \cite{barr_machine_2025}, where the open system dynamics were described using the RC mapping.\par
\begin{figure*}
    \includegraphics[width=0.99\linewidth]{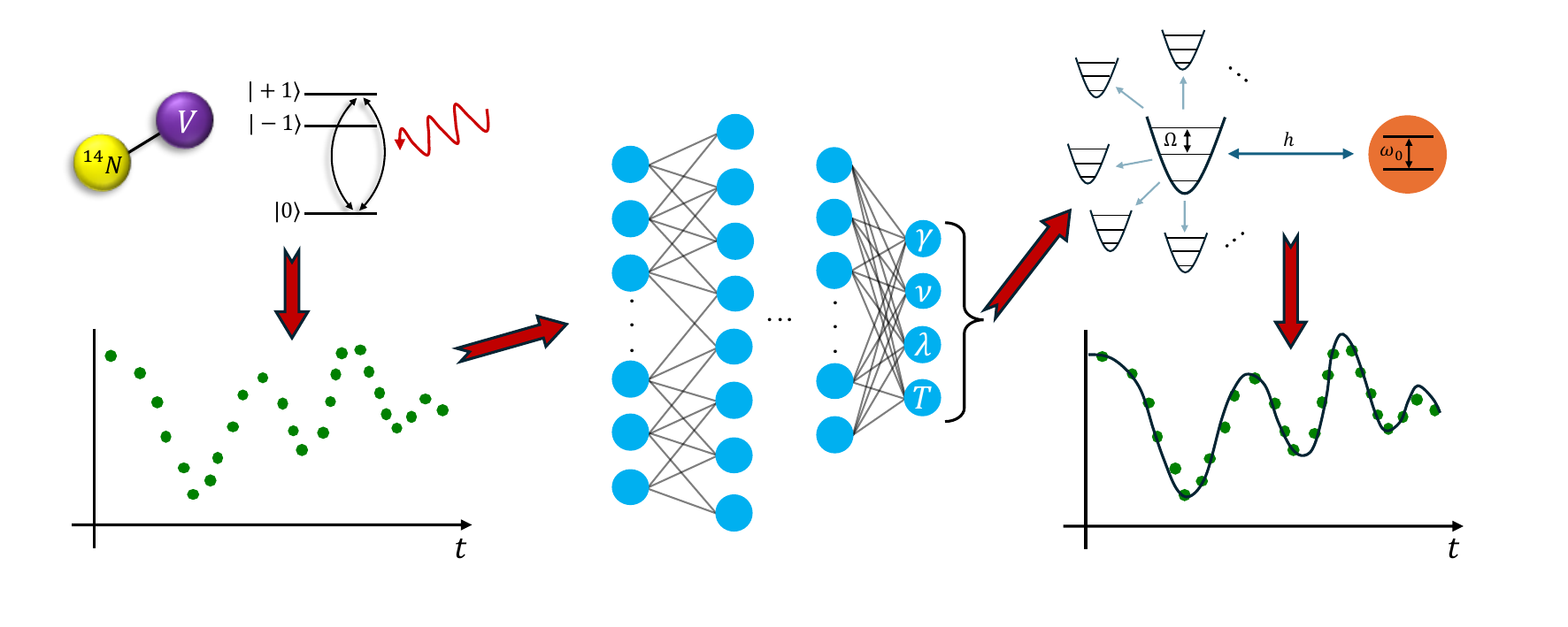}
    \caption{\label{fig:schematic} A schematic of the machine-learning based parameter estimation procedure. A continuous microwave signal drives the spin of the NV centre between two states, resulting in a photoluminescence signal that oscillates and decays (possibly non-monotonically) in time due to coupling to an external environment. This signal is fed into a trained neural network to perform regression on the parameters of the environment. These parameters are used to simulate a reaction coordinate model of the NV dynamics that is able to reproduce the experimental data.}
\end{figure*}
In this work, we model the Rabi dynamics of a nitrogen-vacancy (NV) centre and its environment as a driven spin-boson system, incorporating non-Markovian dynamics via the RC mapping, and use the neural network approach developed in \cite{barr_machine_2025} to infer the parameters of the spectral density. The NV centre has been used for a variety of quantum technology applications, including quantum sensing \cite{schirhagl_nitrogen-vacancy_2014}, quantum computation \cite{zhang_efficient_2020,bartling_universal_2025,dong_experimental_2021}, and quantum thermodynamics \cite{hernandez-gomez_experimental_2020, hernandez-gomez_experimental_2021,hernandez-gomez_autonomous_2022}, as well as for testing other methods of environment characterisation \cite{hernandez-gomez_noise_2018,martina_deep_2023}. It is therefore a good proving ground for new techniques for characterising the environment of open quantum systems. Expanding on existing work that used neural networks trained on noiseless data to estimate a sinlge spectral density parameter, we train neural networks on more realistic, noisy simulated data and obtain all of the spectral density parameters. We find that an effective model can be obtained that reproduces the decay of the NV centre's coherent oscillations within the time range shown to the neural network. A schematic of this method is shown in Fig.~\ref{fig:schematic}. We also evaluate the performance of the neural network approach from a metrological perspective, using the Fisher information and maximum likelihood estimation as benchmarks. Our analysis reveals that the performance of the neural networks in estimating each parameter is correlated with the corresponding Fisher information. Additionally, for certain parameters, the neural network produced estimates with variances on par with maximum likelihood before the asymptotic limit. In light of these findings, we propose that this machine learning approach would fit best in a two-step estimation protocol: the experimental signal would first be fed into the neural network to produce a rough estimate of the parameter values, then maximum likelihood estimation could be applied more efficiently, with the search space narrowed by the neural network estimates.\par
The remainder of this paper is organised as follows. In Sec.~\ref{sec:exp_sys}, we describe the physical system and physical platform at the core of our study. In Section \ref{sec:rc_mapping} we derive a spin-boson model for the NV centre. In Section \ref{sec:nn_approach}, we outline the working principles of the neural network approach used to perform regression on the spectral density parameters of the experimental system. We present the results of modelling the experimental signals and analysis of the neural network performance in Section \ref{sec:results}, and finally, we give the conclusions and outlook in Section \ref{sec:conclusion}.
\section{Experimental System}\label{sec:exp_sys}
\begin{figure}
    \includegraphics[width=0.99\linewidth]{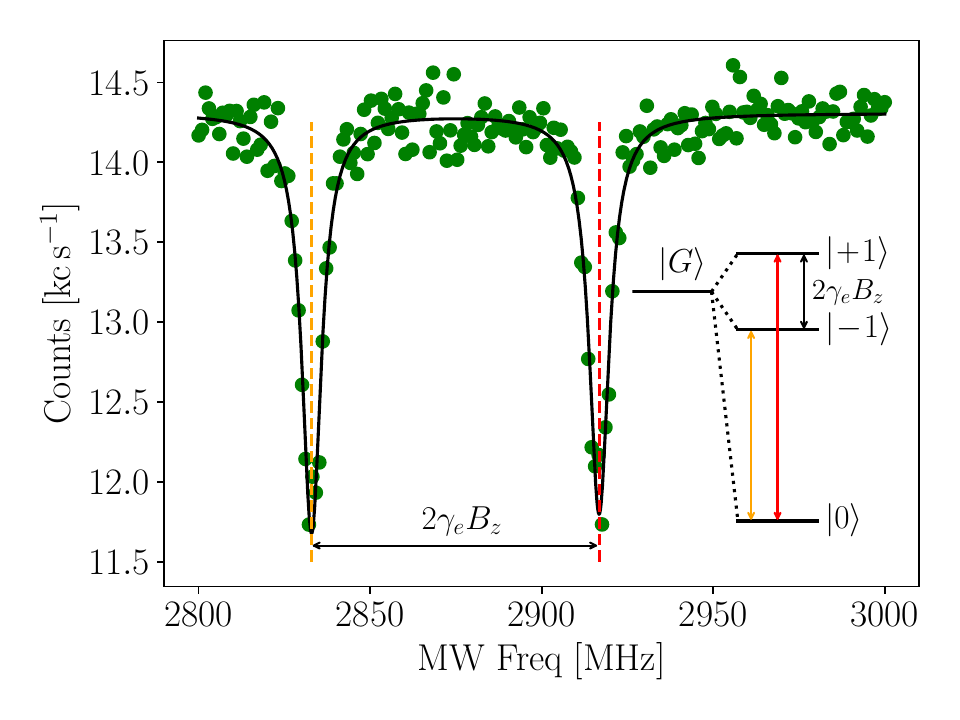}
    \caption{ODMR signal in the ground state of the NV centre. Green points are experimental data, and the black line is a fit to the fine structure Hamiltonian. The relevant energy levels and transitions for this experiment are shown in the inset.}
    \label{fig:odmr}
\end{figure}

The experimental platform at the core of our investigation consists of a single negatively charged nitrogen-vacancy centre in diamond. The NV tip, magnetic field, laser and microwave systems, and photon counting unit are all integrated in the ProteusQ system, a purpose-built quantum sensing device designed by Qnami. The pair of electrons in the NV centre form a spin-1 system whose ground state $\ket{0}$ and degenerate excited states $\ket{\pm1}$ are separated by a $D=2.87\,\mathrm{GHz}$ zero-field splitting. 
The degeneracy between the $\ket{\pm1}$ states can be lifted by applying a static magnetic field along the NV quantisation axis. Taking this to be the $z$-axis, these properties are captured by the ground state Hamiltonian
\begin{equation*}
    H_\mathrm{fine}=D\sigma_{z,e}^2+\gamma_e\vec{B}\!\cdot\!\vec{\sigma}_e,
\end{equation*}
where $\gamma_e=2.8\,\mathrm{MHz}\,\mathrm{G}^{-1}$ is the electron gyromagnetic ratio, $\vec{B}=(B_x,\,B_y,\,B_z)$ is the static magnetic field vector, and $\vec{\sigma}_e=\left(\sigma_{x,e},\,\sigma_{y,e},\,\sigma_{z,e}\right)$ is the vector of spin-1 Pauli matrices in the electronic Hilbert space.
Illuminating the defect with a 532\,nm laser for a short time polarizes the population of the electronic states into $\ket{0}$ and allows for a photoluminescence (PL)-based state readout enabled by the different fluorescence intensities corresponding to the $\ket{0}$ and $\ket{\pm1}$ states. After initialisation into $\ket{0}$, and using a continuous readout laser and microwave drive provided by a nearby antenna, sweeping the microwave frequency allows one to observe the $\ket{0}\leftrightarrow\ket{\pm1}$ transitions, with a dip in PL when the microwave is resonant with each transition, as shown in Fig.~\ref{fig:odmr}. During this optically detected magnetic resonance (ODMR) experiment, a static magnetic field is applied such that the $B_z$ component splits the $\ket{\pm1}$ states by $83.8\,\mathrm{MHz}$. Knowledge of the full vector $\vec{B}$ is unnecessary for the experiments carried out in this work.\par
\begin{figure}
    \includegraphics[width=0.99\linewidth]{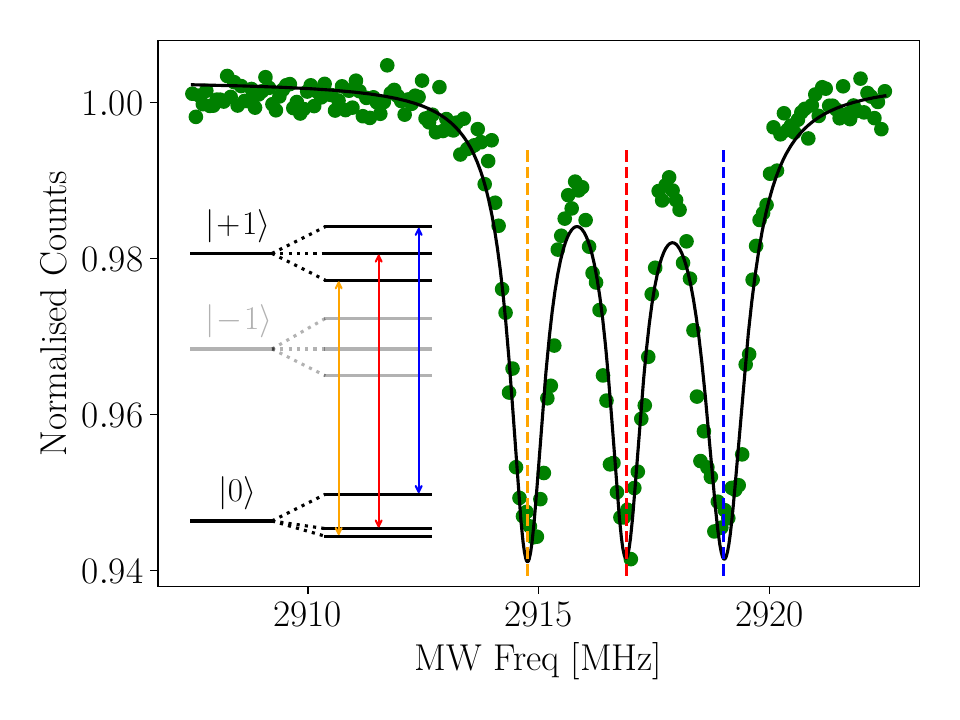}
    \caption{Normalised pulsed ODMR signal in the electronic transition $\ket{0}\rightarrow\ket{+1}$ of the NV centre. Green points are experimental data, and the black line is a fit to the hyperfine Hamiltonian. The relevant energy levels and transitions for this experiment are shown in the inset, and the orange dashed line indicates the transition that is addressed during Rabi oscillation experiments.}
    \label{fig:podmr}
\end{figure}

The coupling of the electronic spin to the nitrogen nuclear spin splits each in a triplet of states that we label as $\ket{j,k}$ ($j,k=-1,0,+1$). Transitions among hyperfine states can be observed using a pulsed ODMR experiment described here. Rabi oscillations are induced between the $\ket{0}$ and $\ket{+1}$ states by applying a continuous, low-power microwave drive resonant with this transition. From these oscillations, one can extract the duration of a $\pi$-pulse -- the time required to fully invert the population from $\ket{0}$ to $\ket{+1}$. The pulsed ODMR experiment is realised by initialising the NV into $\ket{0}$, and applying $\pi$-pulses while sweeping the microwave frequency. In the magnetic field regime used here, the initialization cycle equally populates all hyperfine states in $\ket{0}$ \cite{busaite_dynamic_2020}, resulting in three transitions as shown in Fig.\ \ref{fig:podmr}. This energy level structure is captured by the hyperfine Hamiltonian
\begin{equation}
    H_\mathrm{hyp}=H_\mathrm{fine}+Q{\sigma_{z,I}}^2-\gamma_n\vec{B}\!\cdot\!\vec{\sigma}_I+\vec{\sigma}_e\!\cdot\!\mathbf{A}\!\cdot\!\vec{\sigma}_I,
\end{equation}
where $Q=-4.96\,\mathrm{MHz}$ is the quadrupole splitting, $\gamma_n=31\,\mathrm{kHz}\,\mathrm{G}^{-1}$ is the gyromagnetic ratio of the nitrogen nucleus, $\mathbf{A}=\mathrm{diag}\left(A_\perp,\,A_\perp,\,A_\parallel\right)$ is the hyperfine coupling operator with $A_\perp=-2.7\,\mathrm{MHz}$ and $A_\parallel=-2.14\,\mathrm{MHz}$, and $\vec{\sigma}_I$ is the vector of spin-1 Pauli matrices in the nuclear Hilbert space.
\par
\begin{figure}
    \includegraphics[width=\linewidth]{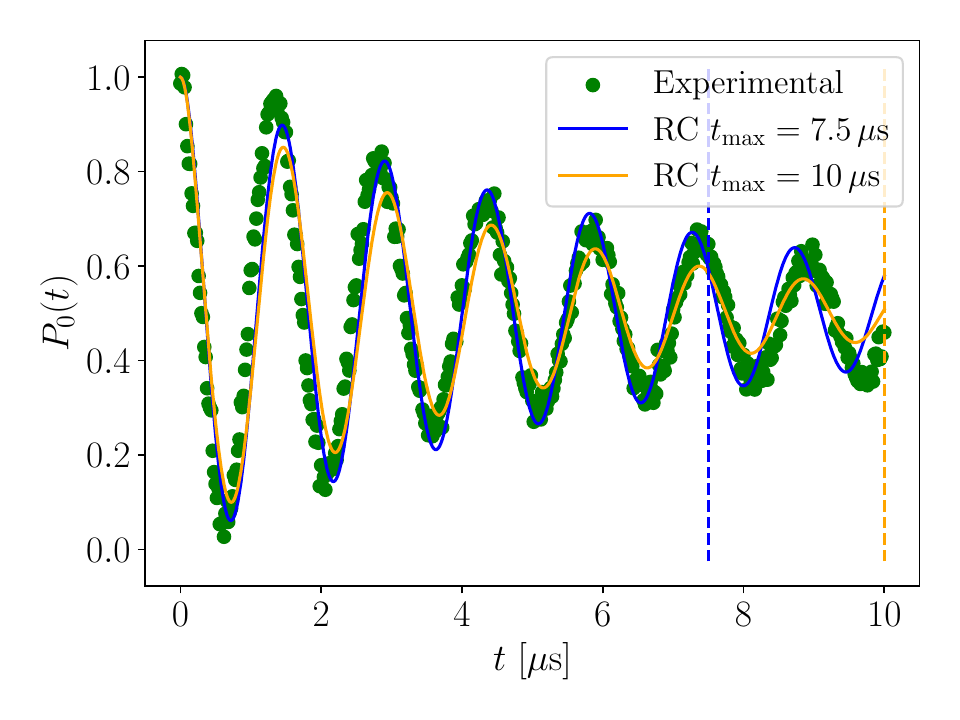}
    \caption{NV centre Rabi oscillations. The green points are experimental data normalised between the PL of the $\ket{0}$ state and the PL after the first $\pi$-pulse. The solid blue (orange) line is the RC model with parameters returned by the neural network trained on simulated data of duration $7.5\,\mu\mathrm{s}$ ($10\,\mu\mathrm{s}$). The vertical dashed lines indicate the duration of experimental data fed into the corresponding neural network.}
    \label{fig:exp_rabi}
\end{figure}

In our Rabi experiments, we initalise the electronic spin into the $\ket{0}$ state and use a low-power microwave drive resonant with the $\ket{0,-1}\leftrightarrow\ket{+1,-1}$ transition, marked by the dashed orange line in Fig.\ \ref{fig:podmr}. The effect of the other hyperfine transitions on the Rabi dynamics will be weak due to the low power of the microwave drive and will be dealt with in Section \ref{sec:rc_mapping}. Under these conditions, we can approximate the NV centre as a two-level system and write the Hamiltonian describing Rabi oscillations as
\begin{equation}
   H_\mathrm{Rabi}=\frac{\omega_0}{2}\sigma_z+\gamma_eB_d\cos{(\omega_dt)}\sigma_x,
\end{equation}
where $\omega_0$ is the energy gap between $\ket{0,-1}$ and $\ket{+1,-1}$, $\sigma_i$ are the spin-$\frac{1}{2}$ Pauli matrices, and $B_d$ and $\omega_d$  the amplitude and frequency of the driving field, respectively. Under the rotating wave approximation in the frame rotating with the driving field, this Hamiltonian becomes
\begin{equation}\label{eqn:rabi}
    H_0=\frac{1}{2}\left(\Omega_R\sigma_x+\Delta\sigma_z\right),
\end{equation}
where $\Omega_R=\gamma_eB_d/\sqrt{2}$ is the Rabi frequency and $\Delta=\omega_d-\omega_0\approx0$ is the detuning. Note that a correction factor of $\sqrt{2}$ is required in the definition of the Rabi frequency to account for the same factor in the spin-1 Pauli matrices. An example of these room-temperature Rabi oscillations is shown in Fig.\ \ref{fig:exp_rabi}. The Fourier transform of this signal allows us to obtain the Rabi frequency as $\Omega_R=0.7\,\mathrm{MHz}$ and hence a driving field strength of $B_d=0.35\,\mathrm{G}$ at the NV, a value that we will retain in the remainder of our analysis.  
\section{Model for the Open System Dynamics}\label{sec:rc_mapping}
At room temperature, a dominant source of noise affecting the NV centre dynamics is the bath of quasilocalised phonons provided by vibrations in the diamond lattice \cite{norambuena_spin-lattice_2018}. The interaction between these two subsystems can be described by the spin-boson model Hamiltonian
\begin{equation}\label{eqn:hamiltonian}
     H=H_0+\sum_i\omega_ib_i^\dagger b_i+X\otimes\sum_ig_i\left(b_i^\dagger +b_i\right),
 \end{equation}
where $H_0$ is the Hamiltonian of the system of interest, $\omega_i$ are the frequencies of the environment modes, $b_i^\dagger$ ($b_i$) is the creation (annihilation) operator of mode $i$, $X$ is the coupling operator in the system Hilbert space, and $g_i$ is the coupling strength of mode $i$ to the system. The distribution of coupling strengths is given by the spectral density function defined as
 \begin{equation}
     J(\omega)=\sum_i|g_i|^2\delta(\omega-\omega_i),
 \end{equation}
where $\delta(\cdot)$ is the Dirac delta function. Assuming that the modes form a continuum, we can write the spectral density function in a way that is consistent with the physical system under consideration. The spectral density of of the quasilocalised phonons can be phenomenologically written as a Lorentzian function peaked at frequency $\nu$, with width $\gamma$ and coupling strength $\lambda$ \cite{norambuena_spin-lattice_2018,norambuena_quantifying_2020},
\begin{equation}\label{eqn:sd}
    J(\omega)=\frac{\lambda^2\gamma\omega}{\left(\omega^2-\nu^2\right)^2+\gamma^2\omega^2}.
\end{equation}
The dynamics of the system resulting from the Hamiltonian in Eq.\ (\ref{eqn:hamiltonian}) can be solved exactly if $\left[H_0,X\right]=0$ \cite{breuer_theory_2007}. Otherwise, the Born and Markov approximations are usually made to derive a Lindblad master equation. Other possible sources of noise include a weak interaction with a spin bath provided by $^{13}\mathrm{C}$ nuclei (1.1\% abundance) and surface charges \cite{kim_decoherence_2015}, and a strong interaction with the native $^{14}\mathrm{N}$ nucleus. The latter of these can provide a source of non-Markovianity, and thus the resulting dynamics cannot be captured using the Born-Markov approximation.\par
To account for non-Markovian dynamics and incorporate the effects of the spin bath, we employ the reaction coordinate mapping. In this formalism, the modes defined by Eq.\ (\ref{eqn:hamiltonian}) are mapped into an effective single mode called a `reaction coordinate' (RC), weakly coupled to a residual bath. The interaction between the RC and the system is treated exactly, allowing for strong interactions and non-Markovian dynamics to be simulated more efficiently. As a result of this mapping, the Hamiltonian (\ref{eqn:hamiltonian}) is transformed into
\begin{equation}\label{eqn:rc_hamiltonian}
    H=H_{S^\prime}+H_{E^\prime}+H_{RCE},
\end{equation}
where
\begin{align}
    H_{S^\prime}=H_0+\Omega b^\dagger b+hX\otimes&\left(b^\dagger+b\right)+\nonumber\\
    &h\frac{X^2}{\Omega}+\sum_{i>1}\frac{l_i^2}{\Omega}\left(b_i^\dagger+b_i\right)^2,
\end{align}
\begin{equation}
    H_{E^\prime}=\sum_{i>1}\Omega_ib_i^\dagger b_i,\quad\mathrm{and}
\end{equation}
\begin{equation}
    H_{RCE}=\left(b^\dagger+b\right)\otimes\sum_{i>1}l_i\left(b_i^\dagger+b_i\right).
\end{equation}
Here, $\Omega$ is the frequency of the RC, $\Omega_i$ is the frequency of residual bath mode $i$, and $b^\dagger$ ($b$) and $b_i^\dagger$ ($b_i$) are their creation (annihilation) operators. The system is no longer coupled directly to the bath, but indirectly via the reaction coordinate, allowing us to more easily simulate the strongly coupled dynamics. The spectral density of the residual bath is related to the original one via
\begin{equation}
    J_\mathrm{RC}(\omega)=\frac{2g_i^2J(\omega)}{\left(\mathcal{P}\int_{-\infty}^\infty\mathrm{d}\omega^\prime\frac{J(\omega)}{(\omega^\prime-\omega)}\right)^2+\pi^2J(\omega)},
\end{equation}
where $\mathcal{P}$ is the Cauchy principle value.\par
With this new Hamiltonian, we can apply the Born and Markov approximations to the coupling between the RC and the residual bath, and assuming the latter is in a thermal state at temperature $T$, the dynamics of the enlarged system (qubit + RC) is governed by the master equation
\begin{widetext}
\begin{equation}
    \frac{\partial\rho_{S^\prime}(t)}{\partial t}=-i\left[H_{S^\prime},\rho_{S^\prime}(t)\right]-\\
    \sum_{i,j}\left(\left[A,A_{ij}\Gamma^+\left(v_{ij}\right)\rho_{S^\prime}(t)\right]+\left[\rho_{S^\prime}(t)A_{ij}\Gamma^-\left(v_{ij}\right),A\right]\right),
\end{equation}
\end{widetext}
where $\rho_{S^\prime}(t)$ is the state of the enlarged stytem, $A=b^\dagger+b$, and $H_{S^\prime}\ket{\phi_i}=\psi_i\ket{\phi_i}$ such that $v_{ij}=\psi_i-\psi_j$. The operator $A_{ij}$ is defined as
\begin{equation}
    A_{ij}=\bra{\phi_i}A\ket{\phi_j}\ket{\phi_i}\bra{\phi_j}.
\end{equation}
Finally, the functions $\Gamma^\pm(\omega)$ are given by
\begin{equation}
\begin{aligned}
    \Gamma^+(\omega)&=\pi
    \begin{cases}
        J_{RC}(\omega)n_B(\omega), & \text{if }\omega>0,\\
        \lim_{\omega\to0}\left[J_{RC}(\omega)n_B(\omega)\right],&\text{if }\omega=0,\\
        J_{RC}(|\omega|)(1+n_B(|\omega|)),&\text{if }\omega<0,
    \end{cases}\\
    \Gamma^-(\omega)&=\pi
    \begin{cases}
        J_{RC}(\omega)(1+n_B(\omega)), & \text{if }\omega>0,\\
        \lim_{\omega\to0}\left[J_{RC}(\omega)(1+n_B(\omega))\right],&\text{if }\omega=0,\\
        J_{RC}(|\omega|)n_B(|\omega|),&\text{if }\omega<0.
    \end{cases}
\end{aligned}
\end{equation}
As we are only interested in the dynamics of system $S$, we trace out the degrees of freedom of the RC. We assume that the initial state of the system and RC is a product state, $\rho_{S^\prime}(0)=\rho_S(0)\otimes\rho_{RC}(0)$, and that
\begin{equation}
    \rho_{RC}(0)=\frac{e^{-\Omega b^\dagger b/T}}{Z_{RC}},\quad Z_{RC}=\mathrm{tr}_{RC}\left(e^{-\Omega b^\dagger b/T}\right).
\end{equation}
We choose the initial state of the system to be $\rho_S(0)=\ket{0}\!\bra{0}$. By examining the Hamiltonian in Eq.~(\ref{eqn:hamiltonian}), it is clear that the open system dynamics are completely described by the system-bath coupling operator, $X$, and the bath spectral density, $J(\omega)$, and temperature, $T$. In turn, $J(\omega)$ is characterized by the three parameters $\gamma$, $\nu$, and $\lambda$.

\section{Neural Network Approach}\label{sec:nn_approach}
Recently, a machine learning-based approach to estimating the central frequency of the spectral density, $\nu$, was proposed~\cite{barr_machine_2025}. In this section, we outline the neural network principles used in this work and extend the analysis to all parameters of the environment, $\left(\gamma,\,\,\nu,\,\lambda,\,T\right)$.\par
To perform regression on these parameters, we use an artificial neural network (NN) composed of layers of neurons, known as a multi-layer perceptron. The first layer receives the data and passes it to the next layer without performing any calculations. In the first hidden layer, each neuron receives a set of inputs, $x_i$, and computes the weighted sum
\begin{equation}
    z=\sum_iw_ix_i+b,
\end{equation}
with weights $w_i$ and a bias $b$. This result is passed to a non-linear activation function $f$ to produce the neuron's output, $y=f(z)$. We use the rectified linear unit function defined as
\begin{equation}
    f(z)=
    \begin{cases}
        z&\text{for }z\geq0,\\
        0&\text{for }z<0.
    \end{cases}
\end{equation}
The output of this layer is passed as an input to the next hidden layer. As the input data propagates through each layer, the outputs become increasingly complex functions of the input data. Finally, the output layer computes the NN's output using a linear activation function $f(z)=z$, with one neuron corresponding to each parameter to be estimated. This choice of output function allows the network to generate unrestricted real-valued outputs. The weights, biases and number of hidden layers are all parameters that are optimised during training.\par
The observable we measure to extract information about the spectral density parameters is $\langle\sigma_z(t)\rangle$ since it is easily accessible experimentally. This time series can be decomposed into its Fourier coefficients, calculated by
\begin{equation}
    Y_k=\sum_{p=0}^{P-1}\langle\sigma_z(t_p)\rangle e^{-2\pi ip/N},
\end{equation}
where $Y_k\in\mathbb{C}$, $P$ is the total number of time steps, $\langle\sigma_z(t_p)\rangle$ is the value of the observable at the $p^\text{th}$ sampled point, and $k=0,\dots,P-1$. Each $Y_k$ is then split into its real and imaginary parts, which are used as inputs into the NN. During training, we use the mean squared error (MSE)  
\begin{equation}
    \mathrm{MSE}=\frac{1}{4N_\text{tr}}\sum_{i=1}^4\sum_{j=1}^{N_\text{tr}}(\eta_{ij}-\hat{\eta}_{ij})^2,
\end{equation}
as a loss function. Here, $N_\text{tr}$ is the number of training examples and $\eta_{ij}$ ($\hat{\eta}_{ij}$) is the true (predicted) value of the $i^\text{th}$ parameter for the $j^\text{th}$ trajectory. Training the NN then reduces to an optimisation problem where the goal is to find the set of network parameters that minimises the loss function. Additionally, we use the $R^2$ metric to track how well the model predicts the true values of each parameter. We thus introduce the parameter
\begin{equation}
    R^2=1-\frac{\text{SS}_\mathrm{res}}{\text{SS}_\mathrm{tot}}
\end{equation}
with
\begin{align}
    \text{SS}_\mathrm{res}&=\sum_{i=1}^4\sum_{j=1}^{N_\mathrm{tr}}(\eta_{ij}-\hat{\eta}_{ij})^2,\\
    \text{SS}_\mathrm{tot}&=\sum_{i=1}^4\sum_{j=1}^{N_\mathrm{tr}}(\eta_{ij}-\langle\eta\rangle)^2,
\end{align}
and where
 $   \langle\eta\rangle=\frac{1}{4N_\mathrm{tr}}\sum_{i=1}^4\sum_{j=1}^{N_\mathrm{tr}}\eta_{ij}$
is the mean of the true parameter values.

To generate training data, we simulate the Rabi dynamics using the Hamiltonian in Eq.~\eqref{eqn:rc_hamiltonian} with $H_0$ defined by Eq.~(\ref{eqn:rabi}), and $X=\sigma_x+\sigma_y+\sigma_z$. This choice of interaction Hamiltonian reflects the hyperfine coupling between the electronic and nuclear spins, and phonons, although other choices may also work. The training dataset is obtained from 6000 sets of environment parameter values randomly sampled from the ranges $\gamma\in\left[0.15,\,0.25\right]\omega_0$, $\nu\in\left[0.54,\,2\right]\omega_0$, $\lambda\in\left[0.1,\,0.25\right]\omega_0$, and $T\in\left[0.3,\,2\right]\omega_0$, and is split in the ratio $\mathrm{train}\!:\!\mathrm{test}\!:\!\mathrm{validate}=4\!:\!1\!:\!1$. The selected parameter value ranges produce a representative set of dynamics.\par
The training procedure for the NNs begins with a hyperparameter search, where a Bayesian search algorithm constructs an NN architecture chosen from a predefined set of options. The hyperparameters tuned by this algorithm are the number of layers, the number of neurons per layer, the batch size, and the learning rate. A starting set is chosen at random, then the training algorithm adjusts the weights, $w_i$, to minimise the MSE over 25 epochs and records the final MSE. The Bayesian prior is then updated to choose the next set of hyperparameter values. This procedure is repeated for a total of 100 sets of hyperparameter values, with the optimal set being the one with the smallest MSE.
\section{Results}\label{sec:results}
\subsection{Modelling Experimental Data}
To obtain parameters that produce an RC model of the NV centre Rabi oscillations in Fig.~\ref{fig:exp_rabi}, we use a neural network trained on simulated data with total duration $t_\mathrm{max}=7.5\,\mu\mathrm{s}$. To make the training signals more realistic, we add noise sampled from a Gaussian distribution with centre zero and width 0.02, which is commensurate with the noise in the experimental data. After a hyperparameter search has been performed, the NN is trained over 1000 epochs. More details of the training data and optimal hyperparameters are given in Appendix \ref{app:nn_training}. Finally, the experimental data is fed into the trained NN, which returns the environment parameters shown in Table \ref{tbl:param_vals}. The corresponding predicted signal is shown in Fig.~\ref{fig:exp_rabi} as a solid blue line. Within the time range shown to the neural network ($t\leq7.5\,\mu\mathrm{s}$), the RC model aligns well with the experimental data. However, allowing the simulation and experiment to continue beyond this range, reveals that the RC model does not accurately predict the experimental dynamics. Despite this, we can feed experimental data of duration  $t_\mathrm{max}=10\,\mu\mathrm{s}$ into a NN trained on data of the same length to obtain a model for this time range. As shown by the solid orange line in Fig.~\ref{fig:exp_rabi}, the RC model again fits well within the time range shown to the network. The Fourier components of the predicted signals are given in Appendix \ref{app:fourier_comps}.\par
\begin{figure}
    \centering
    \includegraphics[width=\linewidth]{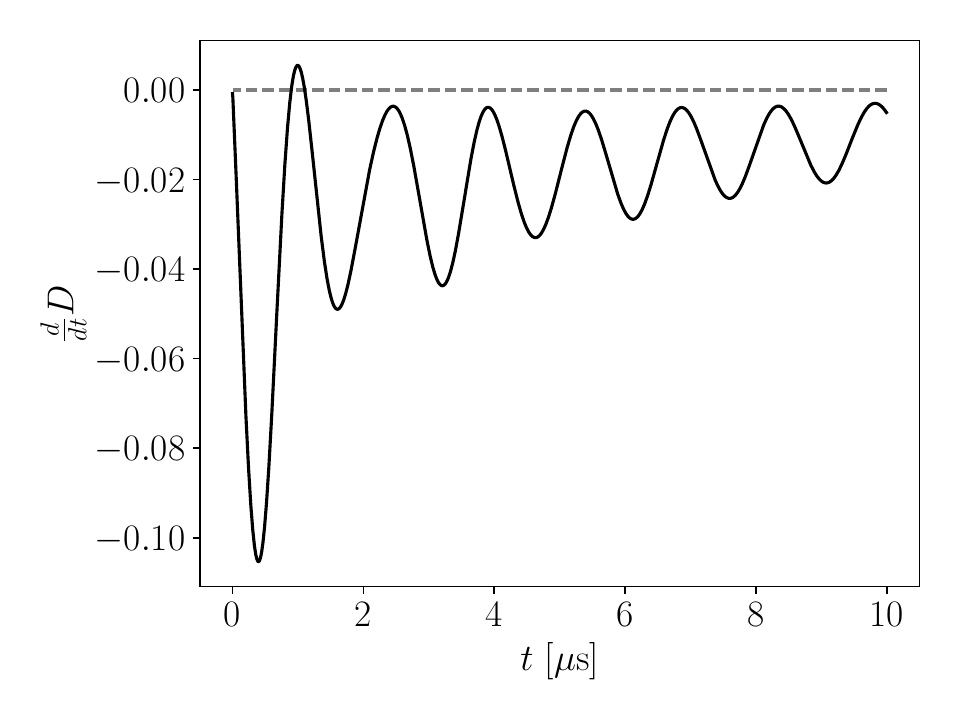}
    \caption{The time derivative of the trace distance between initial states $\ket{0}$ and $\ket{1}$ under the map corresponding to the predicted RC model. Regions above the horizontal dashed line indicate a backflow of information.}
    \label{fig:trace_dist_timederiv}
\end{figure}

\begin{table}
     \begin{tabular}{c|cccc}
     & $\gamma$ & $\nu$ & $\lambda$ & $T$ \\
     \hline\hline
     $t_\mathrm{max}=7.5\mu\mathrm{s}$ & 0.341 & 1.45 & 0.042 & 3.38\\
     \hline
     $t_\mathrm{max}=10\mu\mathrm{s}$ & $0.299$ & $1.43$ & $0.06$ & $2.82$\\
    \end{tabular}
\caption{Parameter values (in units of $\omega_0$) returned by the neural networks.}
\label{tbl:param_vals}
\end{table}
To demonstrate the need to use reaction coordinate mapping to describe these dynamics, we calculate the non-Markovianity of the predicted RC model using the measure based on the trace distance \cite{breuer_measure_2009} defined by
\begin{equation}
    \mathcal{N}=\int_{\Sigma_+}dt\,\frac{d}{dt}{D}\left(\Phi\left[\ket{0}\!\bra{0}\right],\,\Phi\left[\ket{1}\!\bra{1}\right]\right),
\end{equation}
where
 $   D(\rho_1(t),\,\rho_2(t))=\frac{1}{2}\mathrm{Tr}\left[\sqrt{(\rho_1(t)-\rho_2(t))^2}\right]$
is the trace distance between states $\rho_1(t)$ and $\rho_2(t)$, $\Sigma_+$ is the region where $\frac{dD}{dt}>0$, and $\Phi[\cdot]$ is the map corresponding to the predicted RC model. This measure captures non-Markovianity by measuring increases in distinguishability between the two initial states $\ket{0}\!\bra{0}$ and $\ket{1}\!\bra{1}$ when the map $\Phi$ acts on each of them. If $D$ decreases monotonically, this indicates a flow of information into the environment, while if $D$ increases for some times, the two states become more distinguishable, indicating a flow of information from the environment back into the system. For the map corresponding to the $10\,\mu\mathrm{s}$ RC model, the time derivative of $D$ is shown in Fig.\ \ref{fig:trace_dist_timederiv}. Integrating over the regions where $\dot{D}>0$ gives $\mathcal{N}=0.00237$. Hence, the predicted map is (weakly) non-Markovian and the standard Born-Markov approximations will not completely capture the dynamics.\par
These results demonstrate that the NN is capable of finding an effective RC model that accurately represents the NV dynamics within the time domain given. The poor predictive performance of the model is likely due to the physical differences between the actual open dynamics of the NV centre and those represented by the RC model. Specifically, the spin-boson model used here represents a physical model of the phonon bath, but can only produce an effective model of NV dynamics due to the spin baths and coupling to the native nitrogen nucleus.
\subsection{Evaluating the NN Performance}
Benchmarking the performance of NNs for regression purposes is an important aspect of parameter estimation that is often overlooked in the open quantum systems literature. In this section, we investigate how the performance of our NNs changes with the addition of noise into the simulated training set, and explain our observations using the Fisher information. The NN regression performance is then benchmarked against the Cramér-Rao bound (CRB) and maximum likelihood estimation (MLE).\par
\begin{figure*}
 {\bf (a)}\hskip8cm{\bf (b)}\\
    \includegraphics[width=\columnwidth]{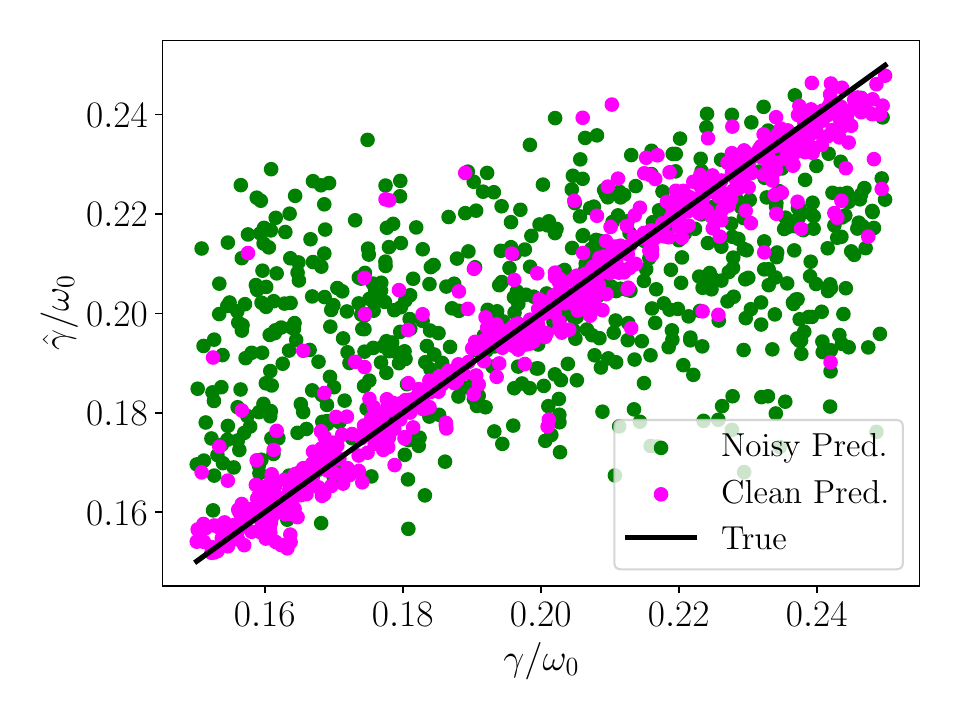}
    \includegraphics[width=\columnwidth]{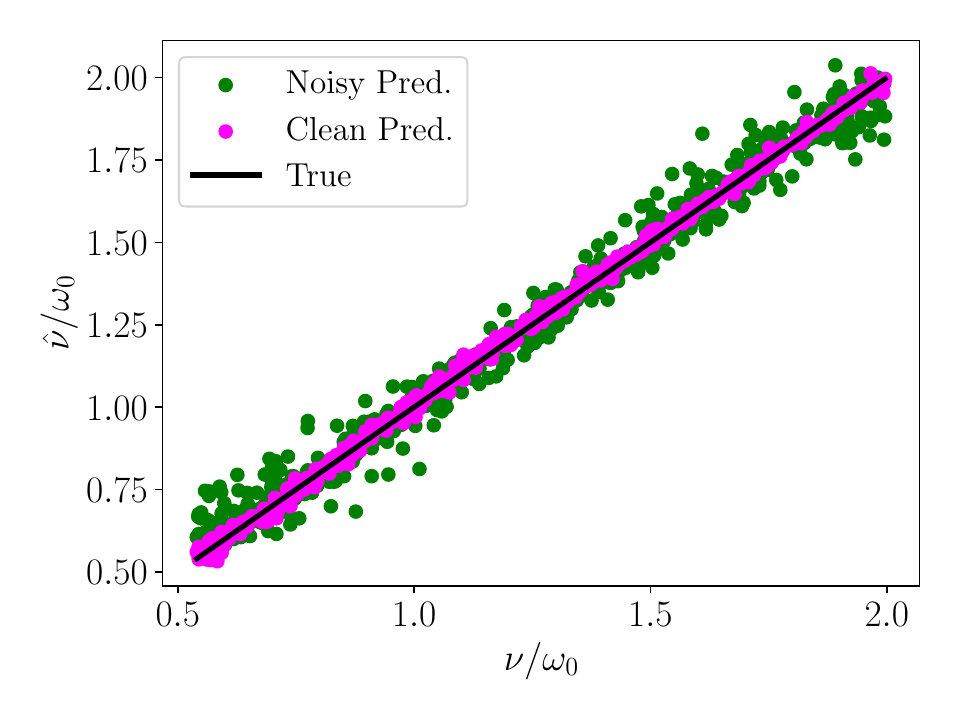}\\
    {\bf (c)}\hskip8cm{\bf (d)}\\
    \includegraphics[width=\columnwidth]{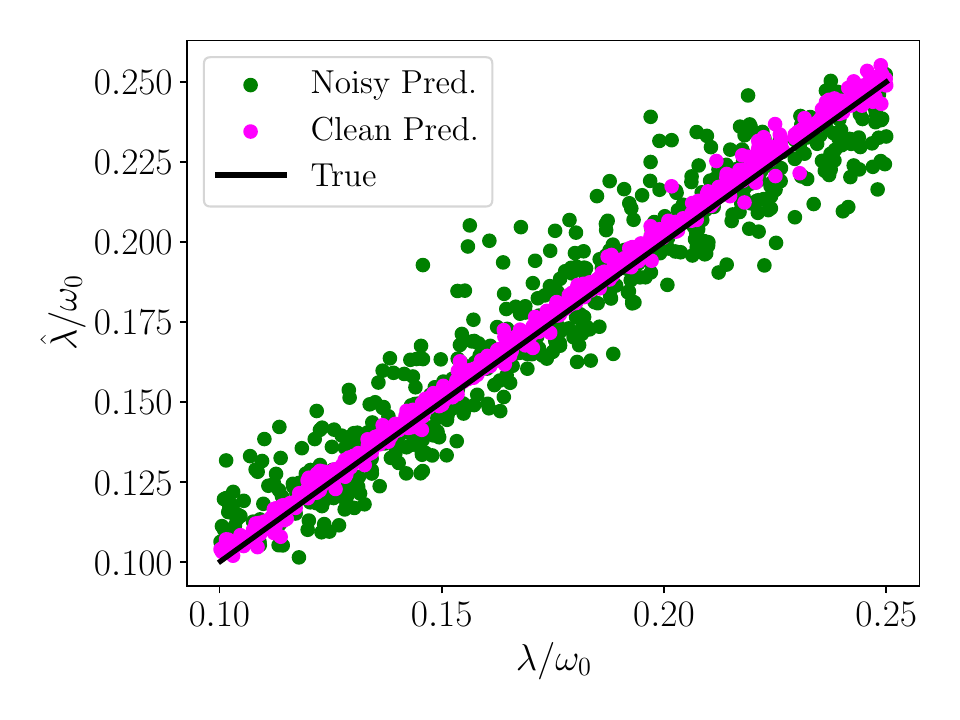}
    \includegraphics[width=\columnwidth]{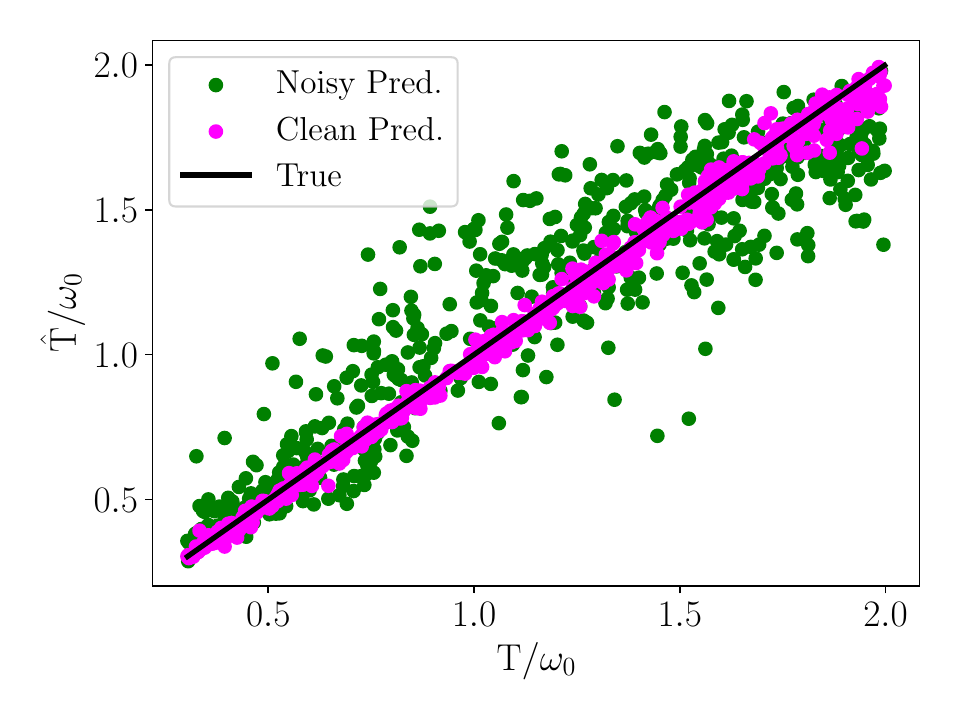}
    \caption{Predicted versus true parameter values for the test set. These predictions were made using the neural networks trained on simulated data with duration $7.5\,\mu\mathrm{s}$, both without (magenta) and with (green) noise added to the training data. The ranges of parameter values used to generate the test set are given in the main text. The line $\hat{\eta_i}=\eta_i$ is also given in each plot, indicating a perfect prediction.}
    \label{fig:pred_vs_actual}
\end{figure*}

We begin by training one NN on clean simulated data and another on simulated data with added artificial noise. As before, the noise is sampled from a normal distribution with centre zero and width 0.02. Once a hyperparameter search has been performed, each NN is trained over 1000 epochs. More details of the training data and optimal hyperparameters are given in Appendix~\ref{app:nn_training}. Finally, to test the predictive ability of the trained NNs, we show each NN the test set and compare the predicted parameter values with the true values, as shown in Fig.~\ref{fig:pred_vs_actual}. On these plots, points located close to the $\hat{\eta}=\eta~(\eta=\gamma,\nu,\lambda,T)$ line indicate that the NN can accurately predict the parameter values. Conversely, points away from this line have been more poorly estimated.

When there is no noise in the training and test sets, the NN predicts the value of each parameter with a high degree of accuracy. However, the predictive performance of the NN trained on data with experimentally realistic noise is worse, as indicated by the larger spread of points. This result is expected as adding noise introduces a precision floor. It is also apparent that both of the NNs are worse at estimating the parameter $\gamma$ than the others. To understand if this is due to the NNs being less suited to estimate $\gamma$ or if $\gamma$ is fundamentally more difficult to estimate, we benchmark the NN performance against the Fisher information and MLE.\par
According to the CRB, the covariance matrix of an unbiased estimator of a vector of parameters $\vec{\eta}$ obeys the matrix inequality~\cite{paris_quantum_2009}
\begin{equation}\label{eqn:CRB_unbiased}
    \mathrm{Cov}(\vec{\eta})\geq\frac{1}{M}F(\vec{\eta})^{-1},
\end{equation}
where $F(\vec{\eta})$ is the Fisher information matrix for the parameter vector $\vec{\eta}$ and POVM $\{\Pi_k\}$, and $M$ is the number of measurements made using this POVM. The elements of the Fisher information matrix are calculated as
\begin{equation}
    F_{ij}(\vec{\eta},t_l)=\sum_k\frac{1}{p(k|\vec{\eta},t_l)}\frac{\partial p(k|\vec{\eta},t_l)}{\partial\eta_i}\frac{\partial p(k|\vec{\eta},t_l)}{\partial\eta_j},
\end{equation}
where $p(k|\vec{\eta},t_l)=\mathrm{Tr}\left[\rho(\vec{\eta},t_l)\Pi_k\right]$ is the probability to obtain outcome $k$ for a given set of parameter values $\vec{\eta}$ at time step $t_l$. As our measurements at different times are independent, the total Fisher information after $N$ time steps is given by
\begin{equation}
    F(\vec{\eta})=\sum_{l=1}^NF(\vec{\eta},t_l).
\end{equation}
The CRB in Eq.~(\ref{eqn:CRB_unbiased}) sets a lower limit on the variance of each parameter and can be saturated in the asymptotic limit ($M\to\infty$) by MLE.\par
For a proper performance evaluation within the Fisher information framework, a few adjustments must be made to the analysis. Firstly, we simulate the dynamics for an environment with a single set of parameter values $\gamma=0.2\omega_0$, $\nu=1.5\omega_0$, $\lambda=0.2\omega_0$, and $T=0.75\omega_0$. These particular values have been chosen so that the NNs interpolate within their training data to produce estimates, and because they result in typical open Rabi dynamics. To ensure consistency in the noise model between estimators, we train an NN on data with projection noise, rather than Gaussian noise. Specifically, we perform $M=750$ projective measurements at each time step using the POVM $\{\ket{0}\!\bra{0},\,\ket{1}\!\bra{1}\}$. To obtain variance estimates for each parameter using both MLE and the NN, we obtain 5000 realisations of the projection noise, giving the same number of estimates for each parameter. To account for the random nature of the training procedure, we train 10 different NNs with the same architecture. A fair comparison between the estimability of different parameters is made by calculating the signal-to-noise ratio for each parameter and method using
\begin{equation}
    \mathrm{SNR}=\frac{\eta_i^2}{\mathrm{Var}(\hat{\eta}_i)},
\end{equation}
For the CRB SNR, we take $\mathrm{Var}(\hat{\eta}_i)=\left(\mathrm{Cov}(\vec{\eta})\right)_{ii}$. Finally, we calculate the bias of each of the NN estimates using
\begin{equation}
    \mathrm{B}(\eta_i)=\langle\hat{\eta}_i\rangle - \eta_i,
\end{equation}
where the average $\langle\hat{\eta}_i\rangle$ is taken over the estimates from a given NN.\par
\begin{figure*}
    \includegraphics[width=\columnwidth]{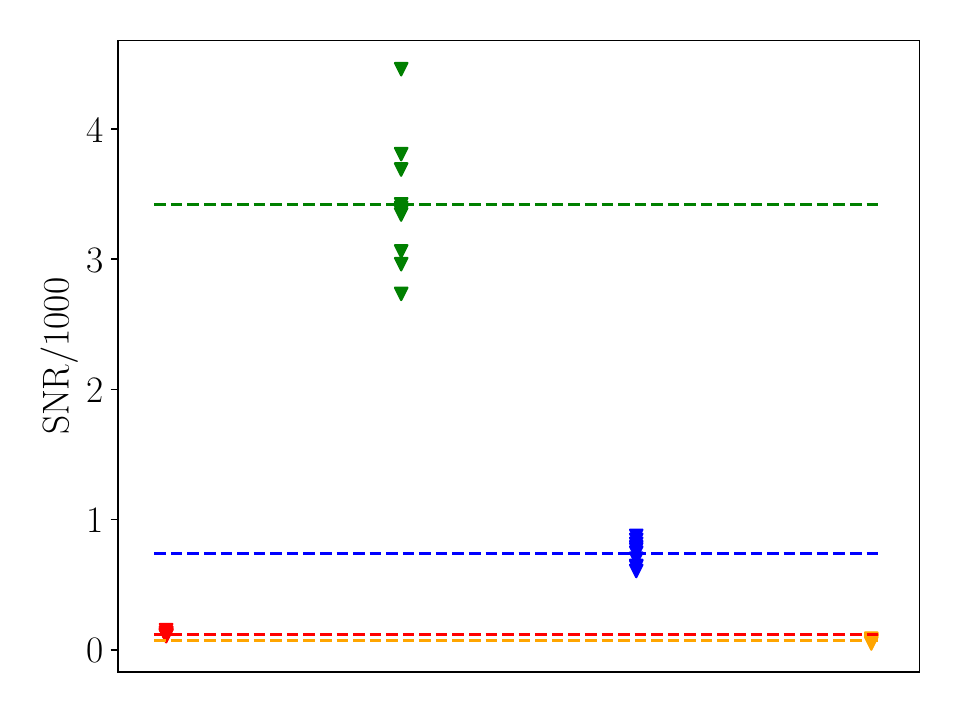}        \includegraphics[width=\columnwidth]{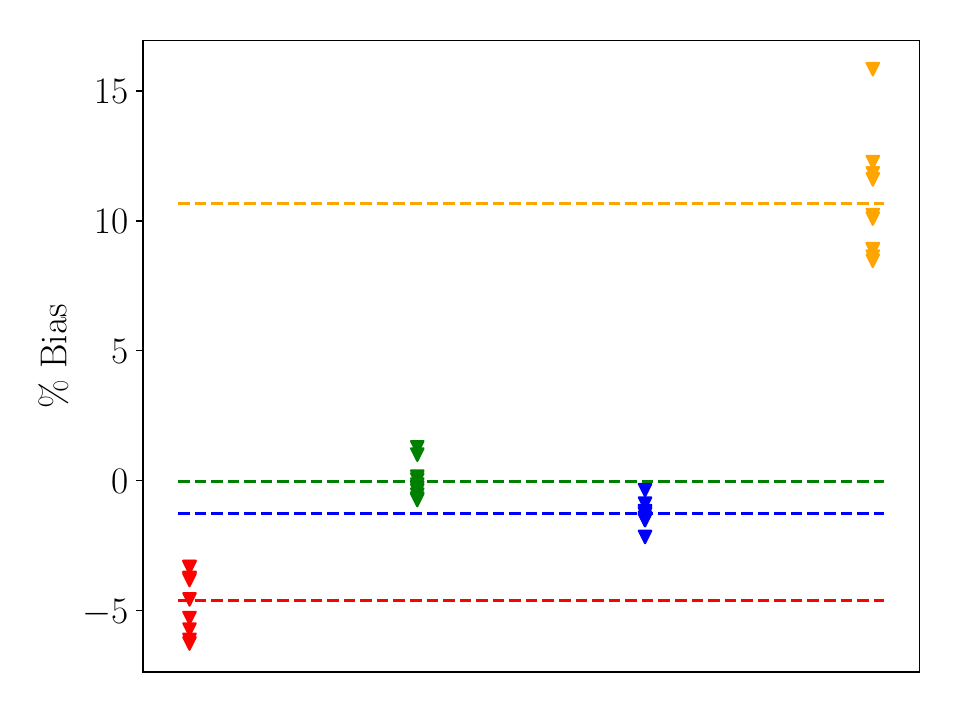}
    \caption{Performance of the neural networks in estimating the parameters $\gamma$ (red), $\nu$ (green), $\lambda$ (blue), and $T$ (orange). For each parameter, there are 10 markers, each corresponding to one of the neural networks, while the dashed lines indicate the mean. On the left is the SNR for each of the parameters, and on the right is their percentage bias.}
    \label{fig:snrs_biases}
\end{figure*}
The results of this analysis are shown in Fig.~\ref{fig:snrs_biases} and Table \ref{tbl:snrs}. According to the SNR calculated from the Fisher information, $\nu$ is the most estimable with an SNR more than ten times larger than the next best, $\lambda$, while $\gamma$ sits at approximately a factor of ten worse again, and temperature trailing slightly behind that. This order of estimability can be understood by considering how each of the parameters affects the dynamics. The central frequency, $\nu$, defines how efficiently the spin and oscillator can coherently exchange energy and phase information. The value chosen here ($\nu=1.5\omega_0$) is relatively close to resonance, resulting in dynamics that are highly sensitive to small changes around this value. The parameter $\lambda$ governs the overall amplitude of the spectral density, which corresponds to the strength of the system-bath interaction. A value of $\lambda=0.2\omega_0$ results in dynamics that are somewhat sensitive to small changes. Finally, the much smaller Fisher information of $\gamma$ and $T$ is likely due to two reasons. The first is that changes in the value of each of these parameters around the values chosen result in much more subtle changes in the dynamics, making estimation more difficult. Secondly, the effects of these parameters on the signal are very similar -- they both contribute to damping the oscillations. The result is that $\gamma$ and $T$ are almost non-identifiable for simultaneous estimation.\par
The estimability ranking is consistent with the average SNR across the 10 neural networks, but the ratios of the neural network SNRs to the Fisher information SNRs are different. For example, the mean NN SNR for $\gamma$ is about $63\%$ of the optimal, while for $\nu$, the average NN SNR is only $15\%$ of the optimal, suggesting that the neural network is prioritising the estimation of some parameters over others. This may be explained by considering that since $\nu$ is relatively easy to estimate according to its CRB SNR, its contribution to the loss function will be small even if the variance of the estimates is far from optimal. This is in contrast to estimates of $\gamma$, whose contribution to the loss function will be very large, even if the NNs are operating close to the CRB. Together, these features mean that the weights are adjusted to prioritise improving $\gamma$ estimates over $\nu$ estimates, with the final result being that the NNs come closer to optimally estimating $\gamma$ than $\nu$. A similar argument applies for the NN performance when estimating $\lambda$ and $T$. We also report the bias $\text{B}(\eta_i)$ as a percentage of the true value of the parameter for each of the neural networks in Fig.~\ref{fig:snrs_biases}. This plot shows that, once averaged over neural networks, the neural network estimator produces unbiased estimates for $\nu$ only, with the remaining parameters biased to varying degrees. This result shows that a comparison to the unbiased CRB in Eq.~\eqref{eqn:CRB_unbiased} will not necessarily reflect the true performance of the neural networks, and that this comparison should be interpreted with care. To capture both the variance and bias, one should use the CRB for biased estimators
\begin{equation}
    \mathrm{Cov}(\vec{\eta})\geq\frac{1}{M}\phi\!\left(\vec{\eta}\right)F\!\left(\vec{\eta}\right)^{-1}\phi\!\left(\vec{\eta}\right)^\mathrm{T},
\end{equation}
where $\phi\left(\vec{\eta}\right)$ is the Jacobian matrix with elements defined by
\begin{equation}
    \phi_{ij}\left(\vec{\eta}\right)=\frac{\partial\left(\mathrm{B}\left(\eta_i\right)+\eta_i\right)}{\partial\eta_j}.
\end{equation}
This analysis, however, is beyond the scope of this work.

The results of this section also show that the variance of the NN estimator is well above the lower bound set by the unbiased CRB. This can be understood by considering that, during training, the NN minimises the loss function simultaneously across the whole parameter range, trading optimal precision in a small region of parameter space for consistent precision across all of parameter space. According to the results in Table \ref{tbl:snrs}, the performance of the NN is comparable to that of an MLE-based approach for estimating both $\gamma$ and $T$, while MLE performs much better for both $\nu$ and $\lambda$, albeit with an SNR still well below the CRB. We also note that the NN approach used here has the advantage that, once trained, the NN requires very few resources to obtain the parameters for a new experimental signal. This is in contrast to MLE, which requires the likelihood function to be maximised every time. Additionally, if the likelihood function is particularly flat, one must impose a narrow search region to allow the algorithm to converge.
\begin{table}[b]
    \begin{tabular}{c|cccc}
     Method & $\gamma$ & $\nu$ & $\lambda$ & $T$ \\
     \hline\hline
     FI & 190 & 23265 & 2278 & 166\\
     \hline
     NN & 120 & 3423 & 742 & 68\\
     \hline
     MLE & 87 & 15896 & 1647 & 92 \\
    \end{tabular}
    \caption{SNRs of each parameter and method. The figures for the NN estimates are the average of the SNRs for each of the 10 NNs.}
    \label{tbl:snrs}
\end{table}
\section{Conclusions}\label{sec:conclusion}
In this work, we have modelled the Rabi dynamics of a single NV centre and its phonon environment using the spin-boson model, and accounted for the interactions with the native \textsuperscript{14}N nucleus and impurity spin baths via the reaction coordinate mapping. Expanding on the work in \cite{barr_machine_2025}, we trained neural networks to identify all parameters of the bath spectral density from noisy experimental data. The RC model identified by the neural network reproduced the experimental data within the estimation window ($t_\mathrm{max}=7.5\,\mu\mathrm{s}$), but failed beyond this time. Nevertheless, a neural network trained on the longer signal ($t_\mathrm{max}=10\,\mu\mathrm{s}$) was able to identify parameter values for an RC model that reproduced the experimental signal over the full time window. These results show that the NNs can identify spectral density parameters of an effective dynamical map, and that a full microscopic model may not be necessary to model systems with complex, non-Markovian environments. This approach substantially reduces the resources needed to simulate such systems, trading predictive power for resource efficiency. Additionally, we expect that the methods developed in this work could be used to identify a unique, predictive model for systems that can be physically described by the spin-boson model + RC mapping, such as ensemble NV-phonon interactions at high temperatures or low NV concentration \cite{norambuena_spin-lattice_2018}.\par
We have also benchmarked the performance of the NNs used for parameter estimation against the Cramér-Rao bound and maximum likelihood estimation. Our findings show that the NNs are not better suited to estimate one parameter over another, but produce estimates with variances that are correlated with those obtained from the Fisher information matrix. While the NN estimator performed well below maximum likelihood for parameters $\nu$\ and $\lambda$, very few resources are required to obtain estimates once the NN has been trained. As such, this approach could be used as the first step in an estimation procedure. The experimental signal would be fed into the NN to reduce the size of the search space, then maximum likelihood estimation could be applied more efficiently to obtain precise estimates.\par
This work opens several avenues for future investigation. First, it would be valuable to push the operating conditions of the experimental platform towards situations exhibiting stronger non-Markovian effects. This may be achieved by polarizing the native nitrogen nuclear spin population to prepare a higher-purity state, the NV centre's interaction with nearby $^{13}\mathrm{C}$ nuclear spins, or via the interaction with a low-temperature phonon bath \cite{norambuena_quantifying_2020}. We expect that the techniques developed in this work could be used to construct effective models of the resulting dynamics, although multiple RCs may be required. A second direction involves the extension of the current analysis to the full parameter space so as to provide a characterization of the metrological capabilities of NN-based estimation approaches. One may also explore how different NN architectures may be able to extract information about the environmental parameters from the experimental signal more efficiently to obtain estimates closer to the CRB.

\begin{acknowledgements}
T.L.\ is grateful to Dr.\ Conor McCluskey for providing training on the experimental apparatus, and to Jessica Barr for fruitful discussions about the machine learning algorithms. T.L.\ and A.K.\ acknowledge financial support from the Engineering and Physical Sciences Research Council (EPSRC) (project reference 2929448). M.P.\ acknowledges financial support from the Royal Society Wolfson Fellowship (RSWF/R3/183013), the Department for the Economy of Northern Ireland under the US-Ireland R\&D Partnership Programme, the PNRR PE Italian National Quantum Science and Technology Institute (PE0000023),  the EU Horizon Europe EIC Pathfinder project QuCoM (GA no.~10032223), and the European Quantum Academy (GA no.~101298535). J.M.G.\ is grateful for funding support from EPSRC (grant number EP/X027074/1) (CAMIE).
\end{acknowledgements}
\appendix

\section{Neural Network Training and Results} \label{app:nn_training}
\begin{figure}
{\bf (a)}\\
        \includegraphics[width=0.99\linewidth]{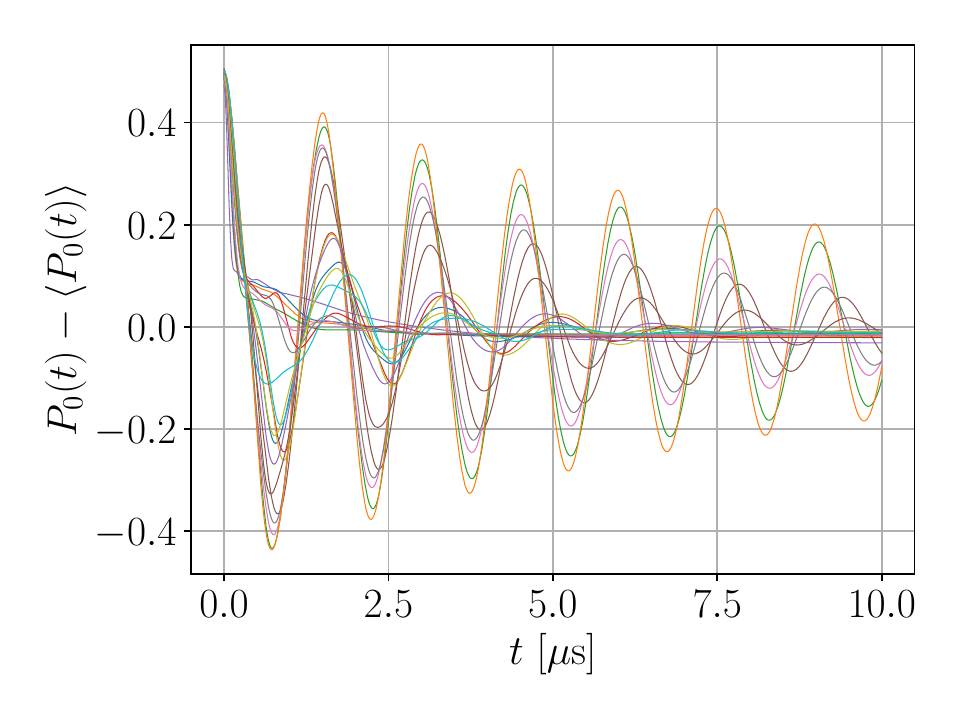}
 {\bf (b)}\\
        \includegraphics[width=0.99\linewidth]{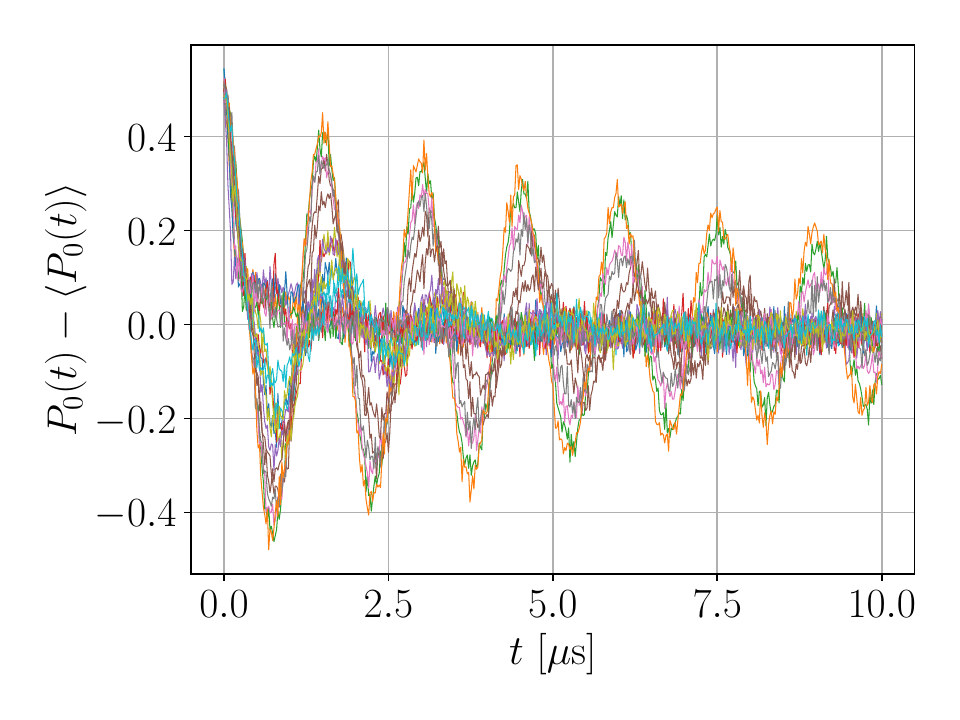}
    \caption{Examples of the simulated Rabi dynamics used for training the NN without noise {\bf (a)} and with noise {\bf (b)}. Each of the trajectories shown was simulated using random spectral density parameter values within the limits given in the main text.}
    \label{fig:training_examples}
\end{figure}

Some examples of the simulated Rabi oscillations used for training the neural networks are shown in Fig.~\ref{fig:training_examples}, both with and without noise, for several sets of randomly selected parameter values within the ranges given in the main text. The mean populations, $\langle P_0(t)\rangle$, have been subtracted to remove the large-amplitude Fourier components at zero frequency.\par
The results of the hyperparameter searches for the neural networks with and without noise in the training data are given in Table \ref{tbl:hyperparams}. For both neural networks trained on noisy data, the noise was sampled from a normal distribution with centre zero and width 0.02.\par
\begin{figure}[H]
\centering
{\bf (a)}\\
        \includegraphics[width=0.99\linewidth]{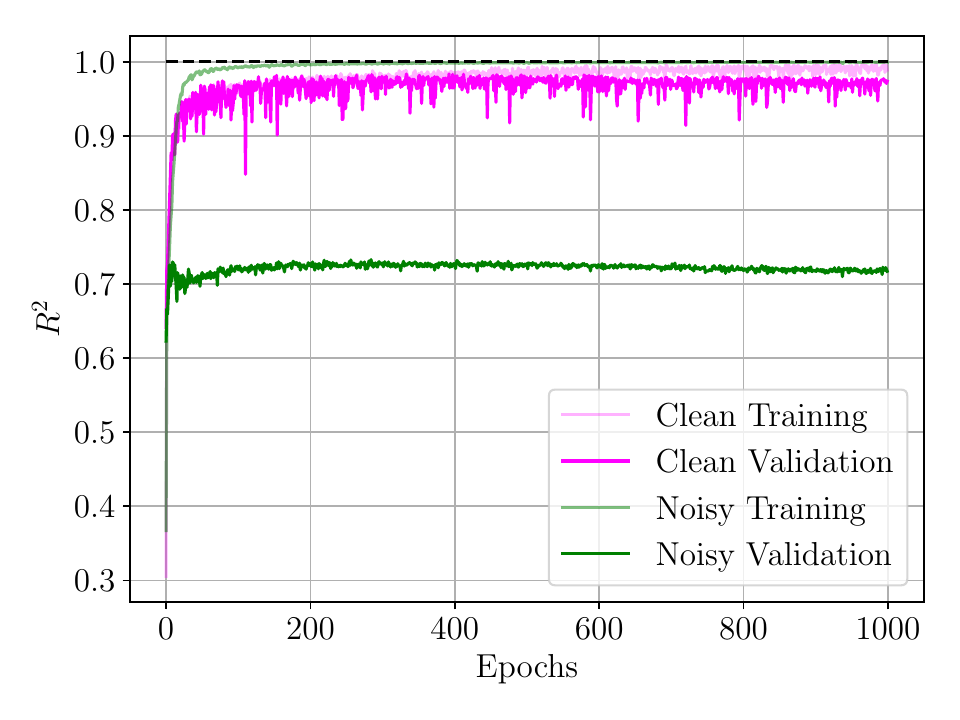}
     {\bf (b)}\\
        \includegraphics[width=0.99\linewidth]{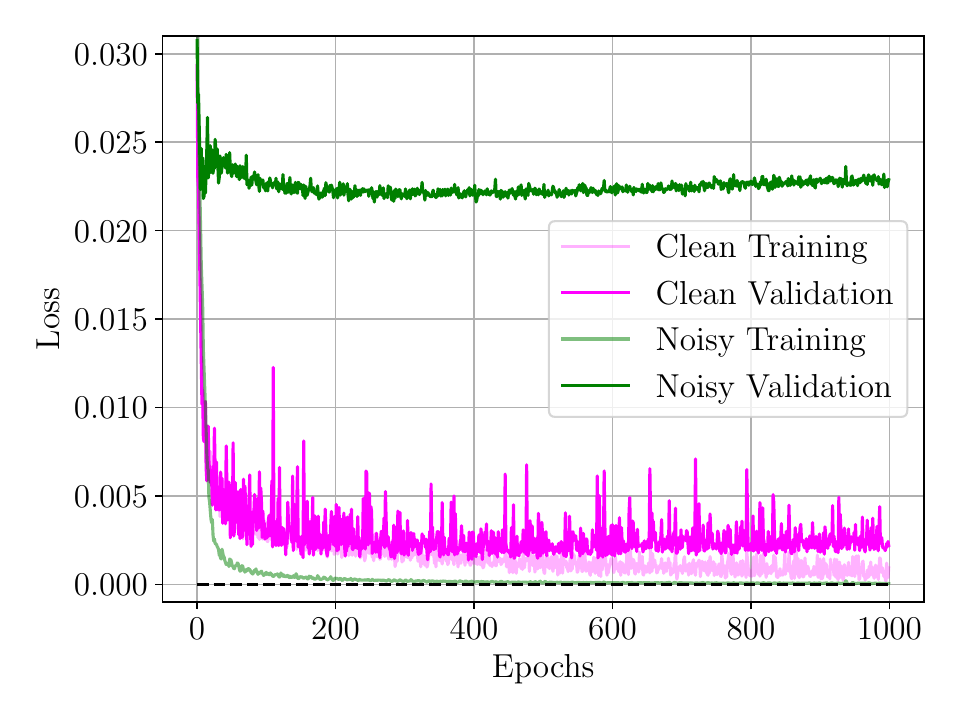}
    \caption{The value of the $R^2$ metric {\bf (a)} and the loss function {\bf (b)} during each training epoch. The magenta and green lines correspond to using training data with no noise and Gaussian noise added, respectively. Light (dark) lines correspond to the metric value with the training (validation) data.}
    \label{fig:nn_performance_metrics}
\end{figure}

At each epoch during the weight training, the loss function given by the MSE is calculated and the weights adjusted to minimise the value of the MSE. Fig.~\ref{fig:nn_performance_metrics} shows the loss function and the value of $R^2$ at each epoch for the training and validation data sets, for the NNs trained on clean and noisy simulated data of duration $7.5\,\mu\mathrm{s}$. Both NNs can memorise the training data set well, as indicated by the convergence of $R^2$ and the loss function to 1 and 0, respectively, when calculated using the training set. However, as expected, the NN trained on noisy data performs more poorly when predicting parameters from unseen data.\par
\section{Fourier Components of Rabi Signals}\label{app:fourier_comps}
\begin{figure}[H]
\centering
  {\bf (a)}\\
    \includegraphics[width=\linewidth]{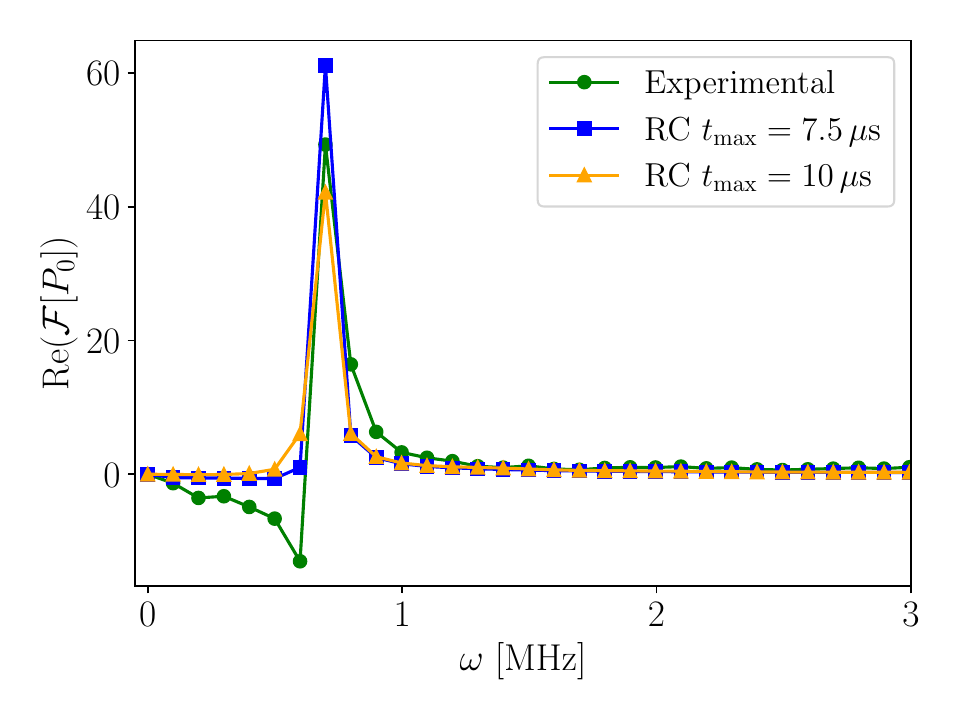}
 {\bf (b)}\\
    \includegraphics[width=\linewidth]{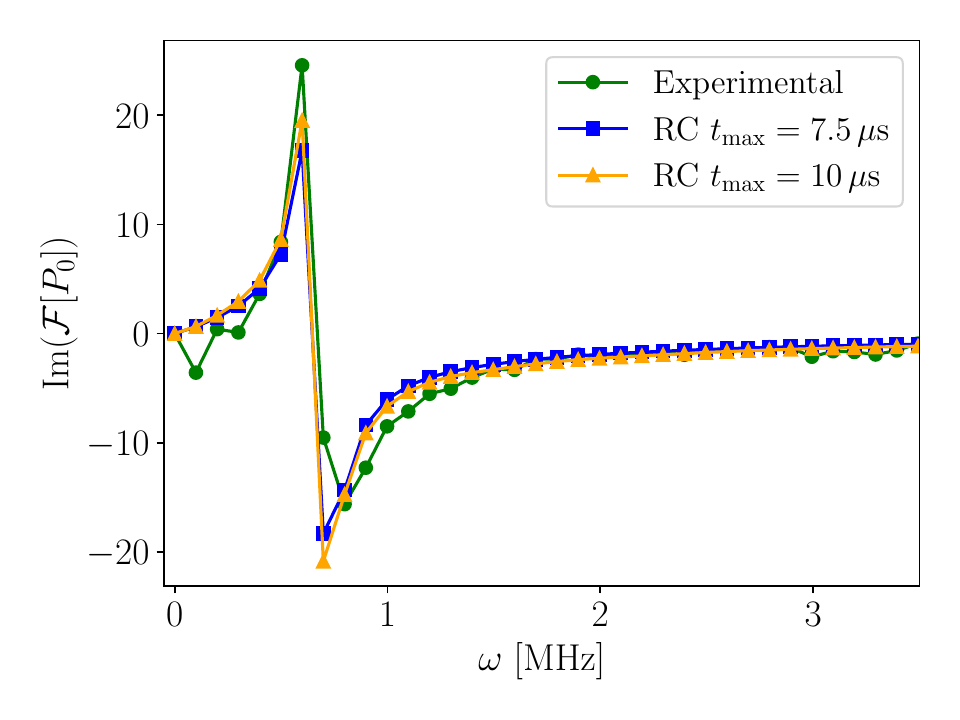}
    \caption{{\bf (a)} Real and {\bf (b)} imaginary parts of the Fourier components of the Rabi signals. The green circles are experimental data and the blue squares (orange triangles) correspond to the RC model with parameters returned by the neural network trained on data of length $7.5\,\mu$s ($10\,\mu$s). The solid lines serve as a guide for the eye.}
    \label{fig:predicted_ft}
\end{figure}

Fig.~\ref{fig:predicted_ft} shows the Fourier components of the signals predicted by NNs trained on data of duration $7.5\,\mu\mathrm{s}$ and $10\,\mu\mathrm{s}$. These correspond to the time traces in Fig.\ \ref{fig:exp_rabi} in the main text.
\bibliography{refs}

\end{document}